\documentclass[preprint,pre,aps,showpacs]{revtex4}
\usepackage{graphicx}
\begin{document}

\title{Bimodality and Coulomb effects with a canonical
thermodynamic model}

\author{G. Chaudhuri \footnote{gargi@physics.mcgill.ca} \footnote{On leave from Variable Energy Cyclotron Centre,
1/AF Bidhan Nagar, Kolkata 700064, India} $^1$, S. Das Gupta$^1$ and F. Gulminelli$^2$}

\affiliation{$^1$Physics Department, McGill University,
Montr{\'e}al, Canada H3A 2T8}

\affiliation{$^2$LPC Caen IN2P3-CNRS/EnsiCaen et Universite, Caen, France}

\date{\today}

\begin{abstract}

The effect of the Coulomb interaction on the phase diagram of finite nuclei is
studied within the Canonical Thermodynamic Model. If Coulomb effects are artificially switched off, this model
shows a phenomenology consistent with the liquid-gas phase transition. The inclusion of Coulomb does not
significantly
affect the phase diagram but it drastically modifies the nature and order parameter of the transition.
A clear understanding of the phenomenon can be achieved looking at the distribution of the largest fragment
produced in each fragmentation event.
Possible connections with experimental observations are outlined.
\end{abstract}


\maketitle

\section{Introduction}
Nuclear multifragmentation in intermediate energy heavy ion collisions
continues to be a topic of intense interest.
In particular it has been recently proposed \cite{pre} that a bimodal
behavior in the distribution of the heaviest fragment produced in
each fragmentation event can be an experimentally measurable
signature of a first-order
transition. Different data-sets have been reported to confirm such
an expectation \cite{pichon,bruno,bonnet} and a lively debate
on its ultimate physical interpretation was
raised\cite{lacroix,trautmann,aichelin,npa}.
To progress on these issues,
here we %study some aspects of this
report on calculations using the canonical thermodynamic model
(CTM) \cite{das1}.
A study of the thermodynamic properties of the model will show that
indeed the bimodality signal can be associated to the finite system
counterpart of a first-order transition
in the framework of our model, but that the nature of such a transition
is deeply modified by the presence of Coulomb with respect to ordinary liquid-gas.

Our ultimate aim is to
confront the model with some data which exist from quasi-projectile (QP)
fragmentation of Au on Au \cite{Neindre,Agostino} but we will
also do, in the beginning,
some calculations for hypothetical nuclei with Coulomb interaction
switched off.  The reason for this preliminary study is that
the role of the Coulomb interaction on the observed phenomenology
can be clearly spotted by such a study.
%some nice features of first order phase transitions emerge which, however, begin to get diluted by Coulomb
interaction.

\section{Basic equations of CTM}

The mathematical machinery for calculating with CTM
has been described before \cite{das1} but to establish our notation
and also list the parameters used in this calculation we need to
go over some details.  The dissociating nucleus breaks up into
clusters.  Each cluster is specified by two indices $a$=number of
nucleons and $j$=number of protons.
The canonical partition function for the fragmenting source with $A$ nucleons and $Z$
protons (neutron number $N=A-Z$) at a given temperature $T$ is given
in our model by
\begin{eqnarray}
Q_{A,Z}=\sum\prod \frac{\omega_{a,j}^{n_{a,j}}}{n_{a,j}!}
\end{eqnarray}
Here the sum is over all possible channels of break-up
which satisfy the conservation laws; $\omega_{a,j}$
is the partition function of one composite with
nucleon number $a$ and proton number $j$ respectively, and $n_{a,j}$ is
the number of this composite in the given channel.
The one-body partition
function $\omega_{a,j}$ is a product of two parts: one arising from
the translational motion of the composite and another from the
intrinsic partition function of the composite:
\begin{eqnarray}
\omega_{a,j}=\frac{V_f}{h^3}(2\pi maT)^{3/2}\times z_{a,j}(int)
\end{eqnarray}
Here $ma$ is the mass of the composite and
$V_f$ is the volume available for translational motion; $V_f$ will
be less than $V$, the volume to which the system has expanded at
break up. We use $V_f = V - V_0$ , where $V_0$ is the normal volume of $A$
nucleons and $Z$ protons.
We will shortly discuss the choice of $z_{a,j}(int)$.  The freeze-out
density in unit of normal nuclear density is $\rho/\rho_0=
V_0/V$.

The probability of a given channel $P(\vec n_{a,j})\equiv P(n_{1,0},
n_{1,1},n_{2,1}......n_{a,j}.......)$ is given by
\begin{eqnarray}
P(\vec n_{a,j})=\frac{1}{Q_{A,Z}}\prod\frac{\omega_{a,j}^{n_{a,j}}}
{n_{a,j}!}
\label{prob_channel}
\end{eqnarray}
The average number of composites with $a$ nucleons and $j$ protons is
seen easily from the above equation to be
\begin{eqnarray}
\langle n_{a,j}\rangle=\omega_{a,j}\frac{Q_{A-a,Z-j}}{Q_{A,Z}}
\end{eqnarray}
The constraints $A=\sum a\times n_{a,j}$ and $Z=\sum j\times n_{a,j}$
can be used to obtain different looking but equivalent recursion relations
for partition functions.  For example
\begin{eqnarray}
Q_{A,Z}=\frac{1}{Z}\sum_{a,j}j\omega_{a,j}Q_{A-a,Z-j}
\end{eqnarray}

We now give the choice of $z_{a,j}(int)$ used in this work.  The
proton and the neutron are fundamental building blocks
thus $z_{1,0}(int)=z_{1,1}(int)=2$
where 2 takes care of the spin degeneracy.  For
deuteron, triton, $^3$He and $^4$He we use $z_{a,j}(int)=(2s_{a,j}+1)\exp(-
\beta e_{a,j}(gr))$ where $\beta=1/T, e_{a,j}(gr)$ is the ground state energy
of the composite and $(2s_{a,j}+1)$ is the experimental spin degeneracy
of the ground state.  Excited states for these very low mass
nuclei are not included.  For mass number $a=5$ and greater we use
the liquid-drop formula.  For nuclei in isolation, this reads
\begin{eqnarray}
z_{a,j}(int)=\exp\frac{1}{T}[W_0a-\sigma(T)a^{2/3}-\kappa\frac{j^2}{a^{1/3}}
-s\frac{(a-2j)^2}{a}+\frac{T^2a}{\epsilon_0}]
\end{eqnarray}
The expression includes the
volume energy, the temperature dependent surface energy, the Coulomb
energy, the symmetry energy and contribution from excited states in the continuum
since the composites are at a non-zero temperature.

In using the thermodynamic model one needs to specify which composites
are allowed in the channels.  For mass number $a$=5 we include proton
numbers 2 and 3.  For mass number $a$=6 we include proton numbers
$z$=2, 3 and 4.  For higher masses we have followed this procedure.
The liquid-drop formula allows one to define drip-lines for a given $a$.
For $a\ge 7$ we include all nuclei within drip-lines.  This choice
allows us to use the same criterion in all cases studied, i.e.,
when the Coulomb is switched off or half turned on or fully
turned on (required to compare with actual data).  We can not prove if
our choice of composites is the best one to use from the point of view of
principles.  But at least it is well defined.  Some study was made in
\cite{Chaudhuri1} on the effects of changing the width of the
ridge of the nuclei
used in computation of properties we seek.  Here we have stuck to
one prescription.

The Coulomb interaction is long-range.  The Coulomb interaction between
different composites can be included in an approximation called
the Wigner-Seitz approximation.  We incorporate this following the
scheme set up in \cite{das1,Bondorf1}.  This requires adding in the argument
of the exponential of Eq.(\ref{eq7}) below a term
$\kappa\frac{j^2}{a^{1/3}}(V_0/V)^{1/3}$.
Defining $R_f^3\equiv\frac{3V}{4\pi}$
the average energy of the system is given by
$\langle E\rangle =\frac{3Z_0^2}{5R_f}+\sum_{i,j}\langle n_{i,j}\rangle e_{i,j}$
where for $a>4$ we have
$e_{a,j}=\frac{3}{2}T+a(-W_0+T^2/\epsilon_0)+\sigma(T)a^{2/3}+
\kappa\frac{j^2}{a^{1/3}}[1.0-(V_0/V)^{1/3}
]+s\frac{(a-2j)^2}
{a^2}-T[\partial\sigma(T)/\partial T]a^{2/3}$.  For $a\leq 4$ we use
$e_{a,j}=\frac{3}{2}T+e_{a,j}(gr)-
\kappa\frac{j^2}{a^{1/3}}(V_0/V)^{1/3}$.  We label
as $\langle E^*\rangle$ the excitation energy: $\langle E^*\rangle=\langle E\rangle-E(gr)$ where $E(gr)$ is
calculated for mass number $A$ and charge $Z$ using the liquid-drop
formula.  The pressure in the model can be shown to be \cite{das1}
simply $p=\frac {T}{V_f}\sum_{a,j}\langle n_{a,j}\rangle$.

\section{The largest and the second largest charge in events}

As recalled in the introduction, the size of the largest fragment
(or equivalently its atomic number $z_1$)
is an especially interesting observable in the multi-fragmentation
problem. Not only it is an experimentally accessible quantity
in exclusive experiments \cite{rivet}, but it is known to provide
an order parameter of fragmentation transitions for a large class
of equilibrium as well as out of equilibrium models\cite{botet}.
The second largest fragment $z_2$, though not so important from the
theoretical viewpoint, has been also extensively used to characterize
the topology of fragmentation\cite{pichon,bellaize,borderie}.

%Experimental data exist on the probability distribution of $z_1$ where $z_1$ is the highest charge of the fragments
in an event.
%Similarly data exist on average values of the second largest charge seen in emerging fragments.  We will denote this
by $z_2$.
Calculations of $z_2$ in CTM have not been done before
and in order to derive the expression, it is advantageous to
derive the one for $z_1$ first.

There is an enormous number of
channels in Eq.(1).  Different channels will have different values
of $z_1$.  For example there is a term $\frac{\omega_{1,0}^N}{N!}
\frac{\omega_{1,1}^Z}{Z!}$.  In this channel the highest value of $Z$
is 1 and thus $z_1$ is 1. The probability
of this channel occurring is (from Eq.(\ref{prob_channel}))
$\frac{1}{Q_{A,Z}}\frac{\omega_{1,0}^N}{N!}\frac{\omega_{1,1}^Z}{Z!}$.
The full partition function can be written as
$Q_{A,Z}=Q_{A,Z}(\omega_{1,0},\omega_{1,1},
\omega_{2,1},\omega_{3,1},\omega_{3,2}........\omega_{a,j}......)$.
If we construct a $Q_{A,Z}$
where we set all $\omega$'s except $\omega_{1,0}$ and $\omega_{1,1}$
to be zero then this
$Q_{A,Z}(\omega_{1,0},\omega_{1,1},0,0,0..............)
=\frac{\omega_{1,0}^N}{N!}\frac{\omega_{1,1}^Z}{Z!}$ and this has
$z_1=1$.  Clearly, $Q_{A,Z}(\omega_{1,0},\omega_{1,1},0,0,0..............)$
is contained in and is a small part of the full partition
function $Q_{A,Z}=Q_{A,Z}(\omega_{1,0},\omega_{1,1},
\omega_{2,1},\omega_{3,1},\omega_{3,2}........\omega_{a,j}......)$.

It will be convenient to introduce a shorthand notation.  Except
when confusion may arise, we will write $\omega_j$ to collectively
mean all of $\omega(a,j)$ where $j$ is fixed but the sum over $a$
runs over the allowed range.
As noted above, the full partition function denoted by $Q_{A,Z}$ is
\begin{eqnarray}
Q_{A,Z}\equiv Q_{A,Z}(\omega_{1,0},\omega_{1,1},\omega_{2,1},\omega_{3,1},
....) \nonumber \\
=Q_{A,Z}(\omega_0,\omega_1,\omega_2,........\omega_Z)
\label{eq7}
\end{eqnarray}

We are now ready to write down a general formula.  Let us ask the question:
what is the probability that a given value $z_m$ occurs as the maximum
charge?  To obtain this we construct a $Q_{Z,N}$ where we set all values of
$\omega_z=0$ when $z>z_m$.  Call this $Q_{A,Z}(z_m)$.  Then
$Q_{A,Z}(z_m)/Q_{A,Z}$ (where $Q_{Z,N}$ is the full partition function with
all the $\omega$'s) is the probability that the maximum charge is
any value between 1 and $z_m$.  Similarly we construct a $Q_{A,Z}(z_m-1)$
where $\omega_z$ is set at zero whenever $z>z_m-1$.  The probability $p_1$
that $z_1$ is $z_m$ is given by
\begin{eqnarray}
p_1(z_1)\equiv p_1(z_m=z_1)=\frac{Q_{A,Z}(z_1)-Q_{A,Z}(z_1-1)}{Q_{A,Z}}
\end{eqnarray}
The average value of $z_1$ is
\begin{eqnarray}
\langle z_1\rangle =\sum z_1 p_1(z_1)
\end{eqnarray}
and the rms of normalised $z_1$ is
\begin{eqnarray}
rms=\sqrt{ \sum p_1(z_1) \left ( (z_1/Z)^2-(\langle z_1\rangle/Z)^2 \right ) }
\end{eqnarray}
We now turn to the calculation of $z_2$. A given charge $z_2$ can be
the second highest charge when (a) there is at least one particle in
$z_2$ but also just one particle in a charge
state $z_1>z_2$ or (b) there is no
no particle with charge $z>z_2$ but $z_2$ occurs at least
twice.  The partition function $Q_a$ for the case (a) is
\begin{eqnarray}
Q_a=\sum \omega_{a,z_1}\times [Q_{A-a,Z-z_1}(z_2)
-Q_{A-a,Z-z_1}(z_2-1)]
\end{eqnarray}
For a fixed $z_1$ the sum over $a$ is over all nuclei within drip-lines.
The sum then goes over all $z_1$'s larger than $z_2$.

For the case (b) we have
\begin{eqnarray}
Q_b=Q_{A,Z}(z_2)-Q_{A,Z}(z_2-1)-\sum_a\omega_{a,z_2}Q_{A-a,Z-z_2}(z_2-1)
\end{eqnarray}
Finally $p_2(z_2)$, the probability that the second largest has the value
$z_2$ is given by
\begin{eqnarray}
p_2(z_2)=[Q_a+Q_b]/Q_{A,Z}
\end{eqnarray}
The average value of $z_2$ results:
%Data exist on $z_2$ where
%
\begin{eqnarray}
\langle z_2\rangle=\sum z_2p_2(z_2)
\end{eqnarray}

\section{Coulomb effects on the fragmentation transition}
%{Equation of State, bimodality, specific heat and effects of the Coulomb force}

\begin{figure}
\includegraphics[width=5.5in,height=4.5in,clip]{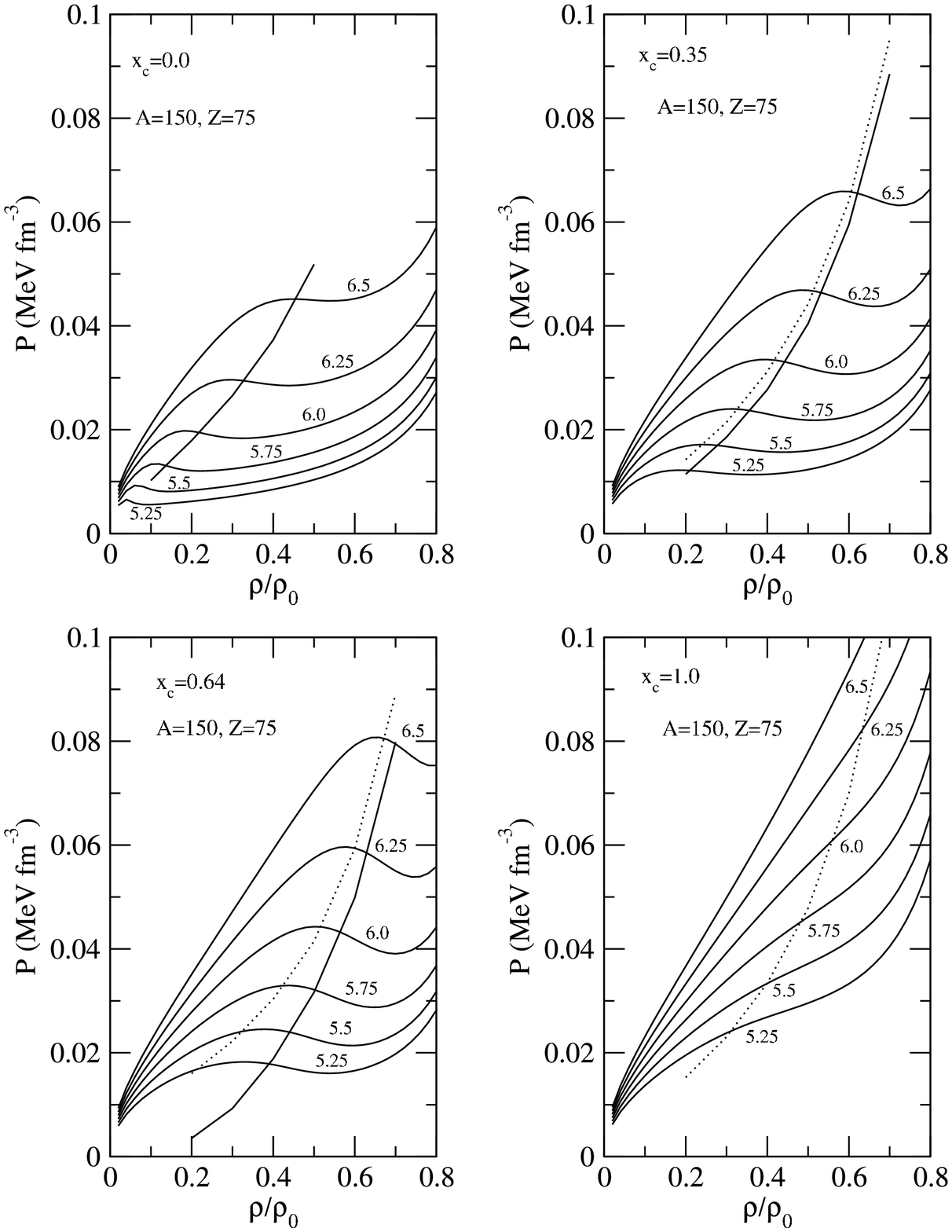}
\caption{ EoS ($p-\rho$) for hypothetical nuclei with a varying
strength of the Coulomb force $x_c$ from zero (upper left) to the
physical value $x_c=1$ (lower right).  The temperatures (in MeV) of
the isothermals are indicated. Along the solid line the two peaks in
$p_1(z_1)$ as a function of $z_1$ are of the same height; $p_1(z_1)$
is the probability that the maximum charge is $z_1$. The dotted line
denotes the line where the specific heat $c_v$ is maximum. The two
lines coincide for $x_c=0$. }\label{fig1}
\end{figure}

For a hypothetical nucleus of mass $A=150$, atomic number $Z=75$ and with no
Coulomb force, we plot the functional
relation between pressure and total density %EoS
in the upper left part of Fig.\ref{fig1}.
A few comments about EoS in the model are in order.

The choice of a symmetric system was done to avoid mixing density
and isospin effects, as we now explain.
It is well known that phase transitions with two conserved charges
(here: proton number $Z$ and neutron $N$ number) can produce fractionation,
i.e. different concentrations in the different phases. Then
to spot the phase diagram of such a system the two densities have to be varied
independently\cite{glendenning,margueron,ducoin}. A projection of
the two-dimensional equation of state on a specific axis,
as for example the $P-\rho$ correlation at fixed $N/Z$,
may be misleading, being continuous even in the occurrence of a
discontinuous (first-order) transition\cite{serot}.
This complication does not arise for symmetric matter,
where fractionation disappears, the direction of phase separation follows
the isoscalar density, and the whole information about the phase diagram
is contained, as for one-fluid systems, in the $P-\rho$ equation of state.
A first-order phase transition in this representation is unambiguously
defined by the presence of a back-bending.
Such behavior is very well known in the framework of the mean-field theory
where it reflects the instability of homogeneous system with respect to phase separation.
There, the appearence of a spinodal region reflects the inadequacy of the model.
One needs to do a Maxwell or Gibbs construction before any correspondence with
experiment can be established.
It should be pointed out that the EoS in models like CTM are very
different from those in the mean-field approximation, although a superficial
look at Fig.\ref{fig1} would suggest this.  There is a spinodal
region in Fig.\ref{fig1} just as there is one in mean-field models,
and in both cases they reflect phase separation.
%However, the backbending in the spinodal region in CTM is miniscule compared to the backbending
%in mean-field models, both in amplitude and in area in the $p-\rho$ plane.
%Two values of $Z$=75 and 60 are chosen.  There is nothing special about these numbers.  Two values of $Z$ are chosen
to see the effects of iso-spin in the model.  The two EoS are quite similar.
%In the literature common figures are EoS in the mean-field model.
However, the backbending in CTM is
an expected feature in the exact evaluation of the thermodynamics of a finite system
which, at the thermodynamic limit, exhibits the discontinuity characteristic
of a first-order phase transition\cite{gross}.
The pressure decrease with density is physical, and it
arises because of finite particle number imposed in the calculation.
For earlier discussions on this see \cite{prl99,Das2}.
In particular, in a simpler version of CTM it was explicitly demonstrated that
the backbending disappears as the system grows and becomes a straight line with zero
slope as expected in a first-order transition\cite{Chaudhuri2}.
The height of the straight line, that is the value of the transition pressure for
a given temperature, is given by the average pressure in the backbending region\cite{Chaudhuri2,noi}.

\begin{figure}
\includegraphics[width=5.5in,height=4.5in,clip]{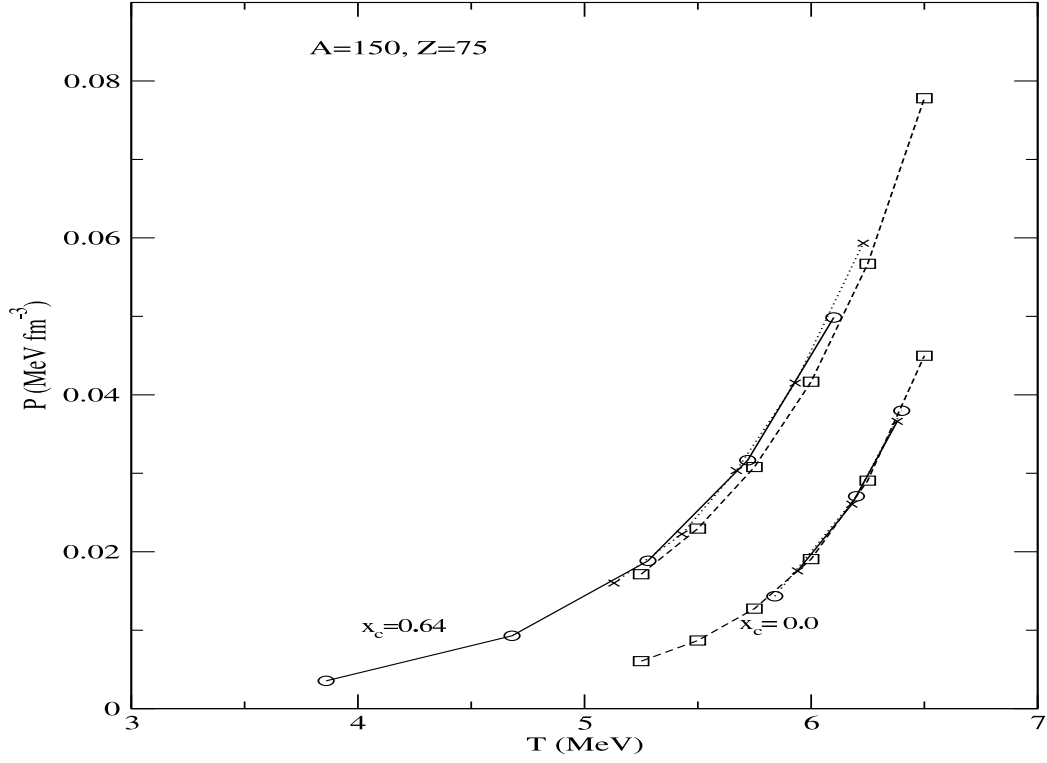}
\caption{Transition pressure as a function of the temperature for an
hypothetical isospin symmetric $A=150,Z=75$ nuclear system
calculated with the three different indicators of the transition:
the average pressure in the backbending region (squares), the curve
along which the two peaks in $p_1(z_1)$ have equal heights
(circles), and the curve along which the $c_v$ is maximum (crosses).
The strength of the Coulomb force is varied from $x_c$=0 to
$x_c=0.64$.}\label{fig2}
\end{figure}

Because of the preceeding discussion, we can conclude from Fig.\ref{fig1} that CTM presents
a first-order transition between a dense, liquid-like phase and a dilute, gas-like one.
The resulting phase diagram is reported in Fig.\ref{fig2}, where the average pressure in the backbending region
has been reported as a function of the temperature.
The qualitative similarity with liquid-gas is apparent.
It has been usual to limit the EoS in the in CTM to $\rho/\rho_0 \le 0.5$.
The argument is that at higher densities the approximation of
replacing the effect of the residual nuclear interaction between composites
by just a constant excluded volume will begin to get worse.  Here however both
in Figs.\ref{fig1}  and \ref{fig2} the curves go beyond this.

We consider now bimodality in the value of $p_1(z_1)$
as a function of $z_1$ where
$p_1(z_1)$ is the probability that the charge of the largest cluster
in an event is $z_1$.  The significance of a bimodal distribution
in a finite system as a signature of first-order transition has
been extensively  discussed in the literature \cite{pre,noi,binder}.
Here we do the calculation
at a  given density $\rho$ for different temperatures to locate the
temperature where the two peaks of the bimodal distribution are nearly
the same height.  Having located the density and temperature we can
then plot this as a point in the $p-\rho$ plane.  At the same density
we can also calculate the specific heat $c_v$ as a function of
temperature and find the temperature where $c_v$ maximises.  This
gives us another point in the $p-\rho$ plane.
Repeating the calculations for different densities  we obtain two curves
in the $p-\rho$ plane.
A maximum in the value of $c_v$ is a generic signature of a phase
transition with finite latent heat rounded by finite size effects.
In the context of nuclear physics, it
was taken as a possible signature of the
transition associated to multifragmentation
a long time ago \cite{Bondorf2}.

The remarkable thing is that without the Coulomb interaction the line representing
bimodality and the maximum of $c_v$ coincide and fall in the
backbending region.
Both curves correspond to maxima of fluctuations: a peak in $c_v=\sigma^2_E/T^2$
means maximal energy fluctuations, and fluctuations in the size of the largest cluster
are maximum at the bimodality point.
In a statistical ensemble where the freeze-out volume would be free to fluctuate under
the effect of an external pressure, it is easy to show that the system would also present
a bimodality in the volume distribution and a maximum in volume fluctuation
at the pressure corresponding to the backbending\cite{noi,raduta}.

The coincidence of all these fluctuation signals agrees with
the expectation that the fragmentation transition is very close to ordinary liquid-gas.
Increasing the available volume at constant temperature, the system experiences
a transition from configurations dominated by a single huge cluster at low excitation
energy, to a highly fragmented pattern at high excitation energy. This transition
can be classified as first-order in the sense that it would become discontinuous
at the thermodynamic limit.
In more technical terms, we can say that energy,
particle density and $z_1$ are all order parameters of the observed transition.
In a system as small as $A=150$ of course all observables are continuous,
however the transition point can be uniquely determined by the location
of the jump (for $\rho$ and $z_1$) or the fluctuation peak (for $E^*$) of the
different order parameters.

This is shown by the phase diagram of Fig.\ref{fig2}. The three different observables
lead to a consistent estimation of the transition line.

The situation changes with Coulomb force.  %This is demonstrated in Fig.2.
We introduce a strength parameter $x_c$ for the Coulomb force: $x_c$=0
means no Coulomb and $x_c=1$ means the full Coulomb.

Figs. \ref{fig1} and \ref{fig2} report the results on the isotherms, fluctuation loci and phase diagram
as the strength of the Coulomb force changes from $x_c=0$ through intermediate values to
its full value $x_c=1$.

Let us look at the $P-\rho$ EoS first.
We can see that the inclusion of the Coulomb interaction results in a shift of
the isotherms, with a gas-like branch appearing at higher density respect to the uncharged case,
and a lower transition temperature for a given pressure.
Moderately charged systems up to $x_c=0.64$ still show a phase diagram
qualitatively similar to the uncharged case, but for the physical system $x_c=1$
the isotherms monotonously grow and the phase transition has disappeared.
These features are in qualitative agreement with previous works\cite{ison,raduta},
as well as with the well-known expectation that the transition temperature should vanish
approaching the drip-lines\cite{bonche}.

The information given by the distribution of the largest fragment closely follows
these thermodynamic findings. The bimodality line falls in the middle of the spinodal region
for all values of $x_c$, and bimodality disappears together with the phase transition
in the fully charged $x_c=1$ case.
Indeed the system $Z=75,A=150$ is beyond the proton drip-line of the model, meaning that huge clusters
close to the size of the source are unbound even at T=0.

The situation is different for the curve of maximum $c_v$.
This curve can be defined even for $x_c=1$, where the system is fragmented at any temperature
and no phase transition is observed. This shows that a peak in $c_v$ is a necessary but not sufficient
condition for a first-order transition\cite{gul99}.
When the transition is present, Fig.\ref{fig2} shows that the $c_v$ observable leads to a
definition of the phase diagram consistent with the isotherms and with the bimodality.
However, we can see from Fig.\ref{fig1} that as the strength of the Coulomb force
changes from $x_c=0$ through intermediate values to
its full value, the splitting between the indicators of phase
transitions progressively increases.
Specifically,  energy fluctuations peak at a density lower than the one one corresponding to the spinodal region in
the charged system. The fluctuation energy peak is thus obtained within a pure phase.

%Similarly the shapes of the bimodality curves for $p_{zmax}$ as a function of $zmax$ changes (Fig.3).
Of course experimentally we only see $x_c$=1. %but the shapes for lower values of $x_c$ are quite interesting
theoretically.
As discussed above, the disappearence of the transition for $x_c$=1 is due to the fact
that the considered system is beyond the proton drip-line.
The more neutron rich system $A=150, Z=60$ has the same size and the same Coulomb energy
as our artificial $A=150, Z=75$ with $x_c=0.64$. This nucleus is not isospin symmetric, meaning that
the $P-\rho$ correlation is not enough to trace the phase diagram of this system and
a complete thermodynamic study would need the analysis of the whole two-dimensional
$P(\rho_n,\rho_p)$ equation of state.

\begin{figure}
\includegraphics[width=5.5in,height=4.5in,clip]{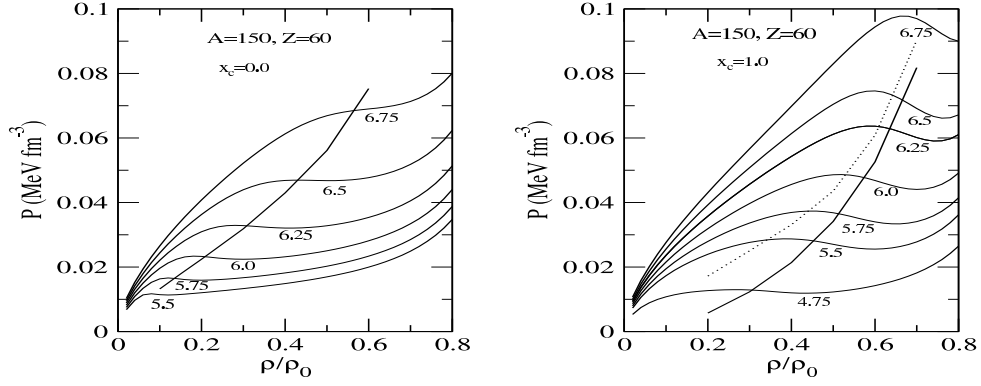}
\caption{ $P-\rho$ correlation for a nuclear system with
$Z=60,A=150$ without ($x_c=0$) and with ($x_c=1$) inclusion of the
Coulomb force.  The temperatures (in MeV) of the isothermals are
indicated. Along the solid line the two peaks in $p_1(z_1)$ as a
function of $z_1$ are of the same height; $p_1(z_1)$ is the
probability that the maximum charge is $z_1$. The dotted line
denotes the line where the specific heat $c_v$ is maximum. The two
lines coincide for $x_c=0$.} \label{fig3}
\end{figure}

However Fig.\ref{fig3} shows that isospin effects have a small influence in
the $P-\rho$ correlation of CTM. This is at variance with mean-field
models for infinite nuclear matter\cite{ducoin,serot}. This is not so surprising
recalling that in CTM the pressure is simply proportional to the total fragment
multiplicity\cite{das1}, and this latter is only slightly affected by the isospin.
Thus the $P-\rho$ isotherms of the asymmetric system $A=150, Z=60$ are very close
to the isotherms of the corresponding symmetric system shown in Fig.\ref{fig1},
once $x_c$ is chosen such as to have the same Coulomb energy as in the physical system.
The same is true for
the bimodality and $c_v$ signal, which are typical isoscalar observables
strongly affected by Coulomb but not very sensitive to isospin.
The preceeding discussion implies that the modification of the phase transition
observed for $0<x_c<1$ in Figs.\ref{fig1},\ref{fig2} may be observed in a physically
existing system like  $A=150, Z=60$. This latter is close to the fragmention sources
studied experimentally in refs.\cite{pichon,bruno,bonnet,Neindre}.

The splitting of the two fluctuation signals observed in Fig.\ref{fig3}
suggests that the nature of the phase transition in charged systems
may be different from that of uncharged ones.
In particular in a transition with non-zero latent heat like liquid-gas
the heat capacity should peak at the transition point, at variance
with our results. This means that energy does not seem a good order
parameter of the transition observed in CTM for charged nuclei.
This last finding is very close to the study of ref.\cite{raduta}.
In that work it was shown that in the MMM fragmentation model
in the isobar canonical ensemble
the two bimodality peaks indicating the phase transition
occur at very different Coulomb energies for heavily charged systems,
but correspond to very similar excitation energies.
If two phases are associated to the same energy, a mixing of the two does not
induce any energy fluctuation, and the energy fluctuation at coexistence
is the same as the energy fluctuation of the pure phases.
Therefore we may expect that in the transition shown by CTM the two phases
should correspond to similar excitation energies, or in other words
that the latent heat of the transition should vanish with Coulomb.

\begin{figure}
\includegraphics[width=5.5in,height=4.5in,clip]{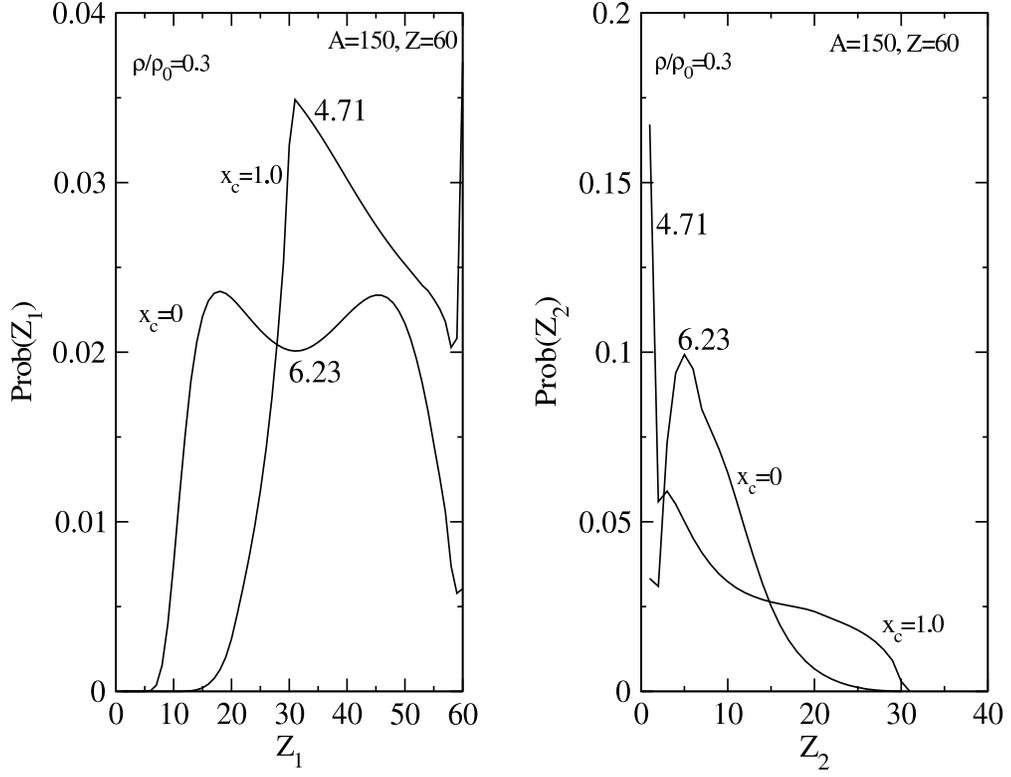}
\caption{ Probability $p_1(z_1)$ as a function of $z_1$ when the
bimodal distribution has nearly same heights for the two peaks. The
system is $A$=150 and $Z=60$.  The strength of the Coulomb force
$x_c$ has been progressively changed from 0 (no Coulomb) to 0.25 to
0.5 to 1 (normal strength).  Note that the shapes of the curves
change very significantly.  The calculation is done at a fixed
density $\rho/\rho_0$=1/3 and the temperature varied till the
bimodal distribution appears.  The temperatures in MeV are
indicated.}\label{fig4}
\end{figure}

To progress in the understanding of the observed phenomenon, we show in Fig.\ref{fig4}
the distribution of the largest $z_1$ and second largest $z_2$
fragment at the bimodality point for the same
system as in Fig.\ref{fig3}. Not only the transition
temperature is lowered by the inclusion of the Coulomb interaction as already observed,
but the shape of the distribution is very different in the presence ($x_c=1$) or absence ($x_c=0$)
of Coulomb.
In the absence of Coulomb the two peaks are similar in shape and close to what is expected
for the liquid-gas phase transition as depicted by the Lattice Gas Model\cite{npa}: a liquid-like
solution with a dominant cluster exhausting about 75\% of the total mass, while the gas-like
solution appearing at the same temperature consists of a much more fragmented configuration
where the largest fragment is about 25\% of the total size, and the second largest is of
comparable size. The situation is completely different for the physical case where Coulomb is accounted
for. In the presence of Coulomb the high-density solution corresponds essentially to
the initial nucleus excited in its internal states. Such configurations give rise to what in nuclear
physics is called an evaporation residue (recall secondary decay is not accounted for in CTM).
The low-density solution is very different from the picture of a nuclear gas: the largest cluster
is peaked at $z_1=Z/2$ and the second largest has a very broad distribution ranging from
$z_2=Z/12$ to $z_2=Z/2$, and is close to the phenomenology expected for hot asymmetric fission.

\section{Summary}

In this paper we have studied the effect of the Coulomb interaction
on the fragmentation transition which is observed at finite temperature
in the Canonical Thermodynamic Model.
The typical behavior which is expected for a finite system counterpart
of the liquid-gas phase transition is only observed if the Coulomb interaction
is artificially switched off.
The transition temperature decreases with increasing Coulomb energy
as observed already in many other models\cite{das1,ison,raduta,bonche}.
More interesting, the fragmentation pattern associated to the transition
changes completely in the presence of Coulomb.
In the spinodal region defined by the backbending of the $p-\rho$
isotherms the distribution of the largest fragment is bimodal.
The two dominant fragmentation patterns defined by the two
peaks at the bimodality point do not correspond to a liquid-gas
phenomenology but are close to a transition from evaporation
to asymmetric fission.

CTM was recently shown to produce results which are close
to another very successful model of nuclear fragmentation, the Copenhagen
SMM\cite{Botvina}. Because of that, we think that the presented results are not
specific to our model but should be characteristic of any model
of fragmentation in statistical equilibrium.

The distribution of the largest fragment as a possible signature
of a fragmentation transition is extensively studied
experimentally\cite{pichon,bruno,bonnet,Neindre}
in quasi-projectile fragmentation of Au+Au collisions.
The comparison of CTM with experimental bimodality data will
allow to progress on the interpretation of the transition
observed in the data which is presently the object
of intense debate\cite{lacroix,trautmann,aichelin,npa}.
This will be the subject of a forthcoming paper.

\section{Acknowledgement}
This work is supported by the Natural Sciences and Engineering Research
Council of Canada and by the National Science Foundation under
Grant No PHY-0606007.

\end{document}